\documentclass{article}

\usepackage{amssymb, amsmath}  
\usepackage[preprint]{neurips_2020}
\usepackage{graphicx,psfrag,epsf}

\usepackage[utf8]{inputenc} 
\usepackage[T1]{fontenc}    
\usepackage{hyperref}       
\usepackage{url}            
\usepackage{booktabs}       
\usepackage{amsfonts}       
\usepackage{nicefrac}       
\usepackage{microtype}      

\newcommand{\V}[1]{\boldsymbol{#1}}  

\DeclareMathOperator*{\CV}{\mathbb{C}ov}

\title{Gaussian Process Model for Estimating Piecewise Continuous Regression Functions}

\author{%
  Chiwoo~Park \\
  Department of Industrial and Manufacturing Engineering\\
  Florida State University\\
  Tallahassee, FL 323212 \\
  \texttt{cpark5@fsu.edu} \\
}

\begin{document}

\maketitle

\begin{abstract}
  This paper presents a Gaussian process (GP) model for estimating piecewise continuous regression functions. In scientific and engineering applications of regression analysis, the underlying regression functions are piecewise continuous in that data follow different continuous regression models for different regions of the data with possible discontinuities between the regions. However, many conventional GP regression approaches are not designed for piecewise regression analysis. We propose a new GP modeling approach for estimating an unknown piecewise continuous regression function. The new GP model seeks for a local GP estimate of an unknown regression function at each test location, using local data neighboring to the test location. To accommodate the possibilities of the local data from different regions, the local data is partitioned into two sides by a local linear boundary, and only the local data belonging to the same side as the test location is used for the regression estimate. This local split works very well when the input regions are bounded by smooth boundaries, so the local linear approximation of the smooth boundaries works well. We estimate the local linear boundary jointly with the other hyperparameters of the GP model, using the maximum likelihood approach. Its computation time is as low as the local GP's time. The superior numerical performance of the proposed approach over the conventional GP modeling approaches is shown using various simulated piecewise regression functions. 
\end{abstract}

\section{Introduction}
In many problems of engineering and science, the response variable of interest suddenly jumps or sharp changes locally. For example, in geostatistics, the responses of different rock strata shows sharp changes at the boundaries of different rock types \citet{kim2005analyzing}. In materials science, the phases of materials suddenly change at the phase boundaries \citet{park2019sequential}. In econometrics, regression discontinuity is often observed in the responses of participants in social programs under different treatments \citep{kang2019errors}. In these examples, there is often little correlation in the response variable values about local jumps or discontinuities. Conventional Gaussian process (GP) models are inappropriate for analyzing those response patterns, because the GP models assume that the response values between two neighboring points are closely correlated \citep{rasmussen2006gaussian}. This paper presents a Gaussian process modeling for estimating discontinuous regression functions. Specifically, we are interested in piecewise continuous functions, i.e. different continuous functions used for modeling different regions of data with possible discontinuities over regional boundaries. 

There are a few studies in Gaussian process literature as to estimation of discontinuous regression functions. The overarching idea of the existing studies is to partition the input domain into multiple regions and fit independent Gaussian process models to the data from different regions. \citet{kim2005analyzing} used the Voronoi tesslation for splitting the domain into regions and fitted independent stationary Gaussian process models to the data from the regions. The number and locations of the regions are jointly estimated with the hyperparameters of the stationary GP models by a Bayesian sampling procedure. \citet{gramacy2008bayesian} used the dyadic treed partitioning for the domain partition and used a stationary Gaussian process model to represent each leaf of the tree. The tree partitioning and leaf Gaussian process models are jointly estimated using the Markov Chain Monte Carlo sampling. Similar treed regression modelings have been used with non-Gaussian process models \citep{malloy2014near, taddy2011dynamic}. These approaches come with multiple limitations. Their partitioning schemes are recursive dyadic splits along axis-aligned direction or triangulation, which are inefficient to partition into regions with complex and curvy regional boundaries, generating many small-sized partitions that contains only a tiny number of training data. They also come with expensive Bayesian sampling approaches to jointly learn the complex domain partition model and local models. 

Regression analysis for piecewise continuous functions has been also studied using non-Gaussian process models. One notable approach is the nonparametric regression approach \citet{qiu2004local, qiu2009jump}. Instead of explicitly estimating the domain partition and piecewise regression models, a local kernel estimate of the unknown regression function is sought using local data. For each test location, the local data neighboring to the test location is considered, and the local data is halved by a hyperplane passing the test point and perpendicular to the local gradient direction at the test location. For each of the two halves, a local kernel estimation is obtained using the local data in the half. Among the two local estimates, the one that gives the smallest residual error is chosen as the final estimate. The local halving serves as a local partition of data into homogeneous regions. However, those non-Bayesian approaches are more appropriate for data densely locating in a low dimensional input space, and they suffer from sparse and high dimensional data. In addition, the approaches does not provide any interval estimates as the GP or other Bayesian models. 

Motivated by the non-parametric regression approach for piecewise regression models, we propose a local Gaussian process model for estimating piecewise continuous regression surfaces. Our GP model seeks for a local GP estimate of an unknown regression function at a test location, using local data neighboring to the test location. The local data is partitioned into pieces, and only a piece of the local data belonging to the same continuous region as the test location is used for the regression estimate. The local partitioning is learned together with other covariance hyperparameters, using the maximum likelihood approach. The proposed approach is the \emph{jump Gaussian process model}. We will describe the new approach in Section \ref{sec:method} and will present our comprehensive numerical studies with various simulated cases in Section \ref{sec:simulation}. Finally, we will conclude in Section \ref{sec:conc}.

\section{Jump Gaussian Process Model} \label{sec:method}
Let $\mathcal{X}$ denote a closed subset of $\mathbb{R}^p$. We consider a problem of estimating an unknown regression function $f: \mathcal{X} \rightarrow \mathbb{R}$ that relates the input $\V{x} \in \mathcal{X}$ to a real response variable. We assume the unknown regression function is piecewise continuous in the form of
\begin{equation} \label{eq:regmodel}
	f(\V{x}) = \sum_{j=1}^J g_j(\V{x}) 1_{A_j}(\V{x}),
\end{equation}
where $A_1, A_2, \ldots, A_J$ are simple connected subsets of $\mathcal{X}$ that partition $\mathcal{X}$ (i.e. disjoint subsets which union forms $\mathcal{X}$), and $g_j(\V{x})$ is a continuous function that represents the regression model on region $A_j$. Here, each function $g_j(\V{x})$ is assumed a realization of Gaussian process with stationary covariance function $c_j(\cdot, \cdot)$ and a  constant mean function $\gamma_j$, and they are mutually independent,
\begin{equation} \label{eq:indep}
	g_j \mbox{ is independent of } g_k \mbox{ for } j \neq k.
\end{equation}
Here, different constant mean functions imply that the function $f$ could have the overall difference between $g_j$ and $g_k$ by the difference of their respective constant mean values, which induces discontinuities in $f$ over the boundary dividing $A_j$ and $A_k$. 

We want to estimate $f$ at a test location $\V{x}_* \in \mathcal{X}$, using $N$ noisy observations of the unknown function, $\mathcal{D} = \{(\V{x}_i, y_i) \in \mathcal{X} \times \mathbb{R}: i = 1, \ldots, N\}$, where
\begin{equation}
	y_i = f(\V{x}_i) + \epsilon_i,
\end{equation}
and independent noises $\epsilon_i \sim \mathcal{N}(0, \sigma^2)$. One plausible approach is to  explicitly estimate $g_j(\V{x})$, $A_j$ and $J$ using the data, like in the Bayesian treed Gaussian process model. Modeling the domain partition requires numerous parameters, and the number of potential domain partitioning ways increases exponentially increases as $J$ increases. The parameter estimation is very complex, involving expensive MCMC calculations for the posterior estimation. 

Here we propose a simpler approach, which takes a local GP estimate of $f$ for each test location $\V{x}_* \in \mathcal{X}$ without explicit estimation of  $g_j(\V{x})$ and $A_j$. We first summarize the conventional local GP approach and discuss its limitation in estimating the piecewise regression function \eqref{eq:regmodel} to explain the needs for a new approach. In the conventional local GP, a small subset of $\mathcal{D}$ nearing a test location $\V{x}_*$ is first chosen. One sensible way for the subset selection is to choose the $n$-nearest neighborhood of $\V{x}_*$. A better approach is to refine the $n$-nearest neighborhood selection by adding or deleting data points based on a selection criterion \citep{gramacy2014local}. Let $I_n(\V{x}_*) \subset \{1, 2, \ldots, N\}$ denote the indices of the selected data so that the local data can be written as
\begin{equation}
	\mathcal{D}_n(\V{x}_*) = \{(\V{x}_i, y_i): i \in I_n(\V{x}_*)\}, 
\end{equation}
and let $\mathcal{X}_n(\V{x}_*)$ denote only the input part of the local data, i.e., 
\begin{equation}
	\mathcal{X}_n(\V{x}_*) = \{\V{x}_i: i \in I_n(\V{x}_*)\}.
\end{equation}
The conventional local GP models the local data as a realization of a Gaussian process with a constant mean function $r(\V{x}) = \gamma $ and covariance function $c(\cdot, \cdot)$. Given the modeling construct, the joint distribution of the local data and the prediction $f(\V{x}_*)$ can be achieved as
\begin{equation} \label{eq:jointpdf_lgp}
	\left[\begin{array}{c}
		f(\V{x}_*) \\
		\V{y}_{n} 
	\end{array}\right] \sim
	\mathcal{N}\left(
	\left[\begin{array}{c}
		0 \\
		\gamma \V{1} 
	\end{array}\right],
	\left[\begin{array}{c c}
		c(\V{x}_*, \V{x}_*) & \V{c}_{n, *}^T\\
		\V{c}_{n, *} & \V{C}_{n},
	\end{array}\right]
	\right)
\end{equation}
where $\V{c}_n = [c(\V{x}_i, \V{x}_*): i \in I_n(\V{x})]$ is a column vector of the covariance values between the local data and the test location, $\V{C}_n = [c(\V{x}_i, \V{x}_k): i, k \in I_n(\V{x}))$ is a square matrix of the covariance values evaluated for all pairs of the local data, and $\V{y}_n = [y_i: i \in I_n(\V{x})]$ is a long column vector containing the response variable values of the local data. By applying the Gaussian conditioning formula to the joint multivariate density, we can obtain the posterior predictive distribution of $f$ at $\V{x}_*$ as the Gaussian distribution with
\begin{equation} \label{eq:localGP}
	\begin{array}{l l}
		\mbox{mean: } 		&   \mu(\V{x}_*) = \V{c}_n^T \V{C}_n^{-1} (\V{y}_n -\gamma \V{1}), \mbox{ and } \\
		\mbox{variance: } 	&   s(\V{x}_*) = c(\V{x}_*, \V{x}_*) - \V{c}_n^T \V{C}_n^{-1} \V{c}_n.
	\end{array}
\end{equation}

\begin{figure}
	\centering
	\includegraphics[width=\textwidth]{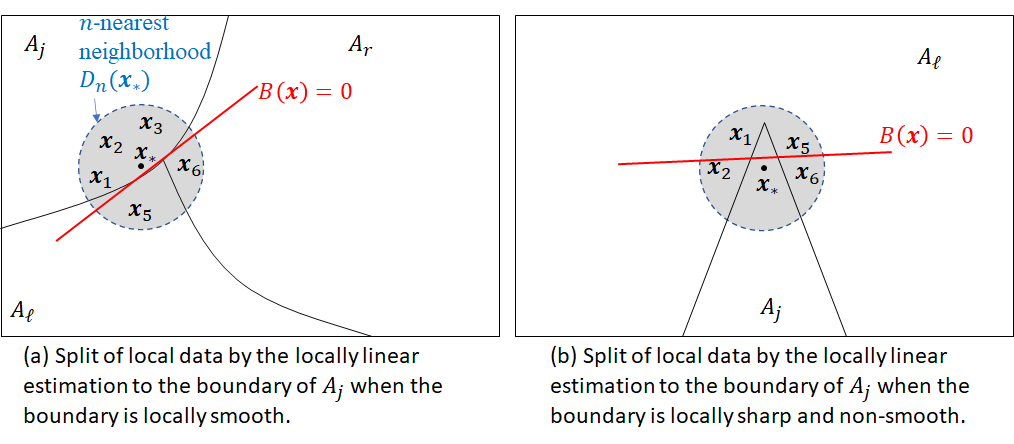}
	\caption{Local Data Split by a Linear Hyperplane $B(\V{x})=0$. For a test location $\V{x}_* \in A_j$, (a) the split successfully separates data from region $A_j$ and data from the other regions if $\V{x}_*$ is near the smooth boundary of $A_j$ as illustrated in figure panel (a), but the split does not work very well when the nearby boundary of $A_j$ as illustrated in figure panel (b) for which a quadratic or higher order polynomial may be needed to represent the local boundary approximation, $B(\V{x}_*)=0$. } \label{fig1}
\end{figure}

This conventional local GP solution \eqref{eq:localGP} would have a huge bias from the $f$ value, when a test location is near the boundary of one of the $J$ regions. Without loss of generality, we assume that $\V{x}_* \in A_j$. When the test location $\V{x}_*$ is interior of the region, all data in $\mathcal{D}_n(\V{x}_*)$ would belong to the region $A_j$, which are all noisy observations of $g_j(\V{x})$, so the conventional local GP solution would work well. However, when $\V{x}_*$ is near or on the boundary of $A_j$, the local data can come from other regions. For example, in Figure \ref{fig1} (a), a test input $\V{x}_*$ belongs to region $A_j$. The local data around the test location contains six data points. Among the six, data points $\V{x}_1$, $\V{x}_2$ and $\V{x}_3$ belongs to $A_j$, but the other data points are outside $A_j$. For that case, if we take the conventional local GP estimate, the estimate would be significantly biased from the ground truth $g_j(\V{x}_*)$, because some of the local data coming from other regions are independent of $g_j$ but influences the estimate of $g_j$.

To mitigate the bias, we propose to divide the local data $\mathcal{D}_n(\V{x}_*)$ into two sides by the boundary of $A_j$, one interior to $A_j$ and another outside $A_j$, and let only the data interior of $A_j$ influence the prediction of $f(\V{x}_*)$. Certainly, this approach is only feasible when we know the boundary of $A_j$, but we do not know the boundary. Instead of estimating the entire boundary, we will estimate a portion of the boundary locally around the test location $\V{x}_*$, using the local data $\mathcal{D}_n(\V{x}_*)$, because the local portion would be sufficient to divide the local data. We will first introduce a parametric model to describe the local boundary and will later explain how the model parameters can be jointly learned with other GP parameters. To introduce the boundary model, please note that when the boundary of $A_j$ is smooth, one can reasonably approximate the boundary locally around the test location $\V{x}_*$ by a linear boundary as illustrated in Figure \ref{fig1} (a), 
\begin{equation}
	B(\V{x}) = [1, \V{x}^T] \V{\beta} = 0.  
\end{equation} 
The approximation would be accurate when the boundary of $A_j$ is smooth, so the first order Taylor approximation of the boundary locally at the boundary point nearest to $\V{x}$ is good enough. Certainly,  the second order or higher-order approximation may be needed when the $A_j$'s boundary is non-smooth and sharp as illustrated in Figure \ref{fig1} (b). For a concise description of the proposed idea, we simply use the linear approximation in this paper. The extension of the idea for the higher order approximation would be not much different from what we are describing here (just more parameters to represent the boundary). 

Now we define a test function $\mathbb{J}$ to determine whether two points $\V{x}$ and $\V{x}'$ belong to the same side of the boundary $B(\V{x}) = 0$,
\begin{equation}
	\mathbb{J}(\V{x}, \V{x}') = \frac{1}{2}\left( 1 + \frac{B(\V{x})}{|B(\V{x})|} \frac{B(\V{x}')}{|B(\V{x}')|} \right),
\end{equation}
which is equal to one if $\V{x}$ and $\V{x}'$ are on the same side of the boundary $B(\V{x})=0$ and is equal to zero otherwise. Based on that, we can divide $I_n(\V{x}_*)$ into two parts: $I_{n,*}(\V{x}_*) = \{ i \in I_n(\V{x}_*): J(\V{x}_i, \V{x}_*) = 1\}$ and $I_{n,o}(\V{x}_*) = \{ i \in I_n(\V{x}_*): J(\V{x}_i, \V{x}_*) = 0 \}$. Accordingly, the local data are divided to
\begin{equation}
	\begin{split}
		& \mathcal{D}_{n,*}(\V{x}_*) = \{ (\V{x}_i, y_i):  i \in I_{n,*}(\V{x}_*)\} \\
		& \mathcal{D}_{n,o}(\V{x}_*) = \{ (\V{x}_i, y_i):  i \in I_{n,o}(\V{x}_*)\}.
	\end{split}
\end{equation}
The data in $\mathcal{D}_{n,*}(\V{x}_*)$ are on the same side of $B(\V{x})=0$ as the test location, and the data in $\mathcal{D}_{n,o}(\V{x}_*)$ are on the other side. Accordingly, we split $\V{y}_n$ into $\V{y}_{n,*} = [y_i: i \in I_{n,*}(\V{x}_*)]$ and $\V{y}_{n,o} = [y_i: i \in I_{n,o}(\V{x}_o)]$.

If the boundary $B(\V{x})=0$ locally approximates the boundary of $A_j$ well as in Figure \ref{fig1} (a), the data $\mathcal{D}_{n,*}(\V{x}_*)$ would be mostly within $A_j$, which are highly correlated to $g_j$, while the data $\mathcal{D}_{n,o}(\V{x}_*)$ would be mostly outside $A_j$, which should be independent of $g_j$ due to model independence \eqref{eq:indep}. We model the two sides of the local data as independent Gaussian process models: Gaussian process with a constant mean $\gamma_{*}$ and a covariance function $c_*(\cdot, \cdot)$ for $\mathcal{D}_{n,*}(\V{x}_*)$ and a Gaussian process with a constant mean $\gamma_{o}$ and a covariance function $c_o(\cdot, \cdot)$ for $\mathcal{D}_{n,o}(\V{x}_*)$. These two independent GP models are represented as a single GP model with the mean function
\begin{equation*}
	r_{\mathbb{J}}(\V{x}) = \mathbb{J}(\V{x}, \V{x}_*) \gamma_*  + (1-\mathbb{J}(\V{x}, \V{x}_*))\gamma_o,
\end{equation*}
and the covariance function 
\begin{equation}
	c_{\mathbb{J}}(\V{x}, \V{x}') = c_*(\V{x}, \V{x}') \mathbb{J}(\V{x}, \V{x}') + c_o(\V{x}, \V{x}') \mathbb{J}(\V{x}, \V{x}'). 
\end{equation}
One can easily show that $c_{\mathbb{J}}(\V{x}, \V{x}')$ is a valid covariance function, because every square covariance matrix evaluated with $c_{\mathbb{J}}$ is a block diagonal matrix with each block as a square covariance matrix of $c_*$ or $c_o$, which is positive definite. The new covariance function satisfies the following condition: 
\begin{equation}
	c_{\mathbb{J}}(\V{x}, \V{x}') = \left\{ \begin{array}{l l}
		c_*(\V{x}, \V{x}') & \mbox{ if both } \V{x}, \V{x}' \mbox{ are the same side of the boundary as } \V{x}_* \\
		c_o(\V{x}, \V{x}') & \mbox{ if both } \V{x}, \V{x}' \mbox{ are the different side of the boundary from } \V{x}_* \\
		0 & \mbox{ otherwise. }
	\end{array}\right.
\end{equation}
Please note that the covariance $c_{\mathbb{J}}$ between a data point in $\mathcal{D}_{n,*}(\V{x}_*)$ and a data point $\mathcal{D}_{n,o}(\V{x}_*)$ is zero. So, $\mathcal{D}_{n,*}(\V{x}_*)$ and $\mathcal{D}_{n,o}(\V{x}_*)$ are independent. 

Under the model, the unknown quantity $f(\V{x}_*)$ is uncorrelated to $\V{y}_{n,o}$ and is correlated to $\V{y}_{n, *}$ with
\begin{equation*}
	\CV[f(\V{x}_*), y_i] = c_{\mathbb{J}}( \V{x}_*, \V{x}_i) = c_*(\V{x}_*, \V{x}_i) \quad \forall \quad i \in I_{n,*}(\V{x}_*).
\end{equation*}
Let  $\V{c}_{\mathbb{J}, n, *} = [c_{\mathbb{J}}(\V{x}_i, \V{x}_*): i \in I_{n,*}(\V{x})]$ represent a column vector of the covariance values. The two part, $\V{y}_{n,*}$ and $\V{y}_{n,o}$, are uncorrelated, based on the covariance model $c_{\mathbb{J}}$. The data points in $\V{y}_{n,*}$ are mutually correlated in that for each $i, k \in I_{n, *}(\V{x}_*)$, 
\begin{equation*}
	\CV[y_i, y_k] = c_{\mathbb{J}}( \V{x}_i, \V{x}_k) = c_*(\V{x}_i, \V{x}_k).
\end{equation*}
Let $\V{C}_{\mathbb{J},n, *} = [c_{\mathbb{J}}(\V{x}_i, \V{x}_k): i, k \in I_{n,*}(\V{x}))$ represent a square matrix of the covariance values evaluated for all pairs of $\mathcal{D}_{n,*}(\V{x})$. Similarly, the data points in $\V{y}_{n,o}$ are correlated. Let $\V{C}_{\mathbb{J},n, o} = [c_{\mathbb{J}}(\V{x}_i, \V{x}_k): i, k \in I_{n,o}(\V{x})]$ represent a square matrix of the covariance values evaluated for all pairs of $\mathcal{D}_{n,o}(\V{x})$. The joint distribution of $f(\V{x}_*)$, $\V{y}_{n,*}$ and $\V{y}_{n,o}$ is a multivariate normal distribution,
\begin{equation} \label{eq:jointpdf}
	\left[\begin{array}{c}
		f(\V{x}_*) \\
		\V{y}_{n,*} \\
		\V{y}_{n,o}  
	\end{array}\right] \sim
	\mathcal{N}\left(
	\left[\begin{array}{c}
		0 \\
		\gamma_* \V{1} \\
		\gamma_o \V{1}  
	\end{array}\right],
	\left[\begin{array}{c c c}
		c_{\mathbb{J}}(\V{x}_*, \V{x}_*) & \V{c}_{\mathbb{J},n, *}^T & \V{0}\\
		\V{c}_{\mathbb{J},n, *} & \V{C}_{\mathbb{J},n, *} & \V{0} \\
		\V{0} & \V{0} & \V{C}_{\mathbb{J},n, o}
	\end{array}\right]
	\right).
\end{equation}
By applying the Gaussian conditioning formula to the joint density, we can achieve the posterior distribution of $f(\V{x}_*)$, which is a normal distribution with 
\begin{equation} \label{eq:localGP3}
	\begin{array}{l l}
		\mbox{mean: } 		&   \mu_{\mathbb{J}}(\V{x}_*) = \V{c}_{\mathbb{J},n, *}^T \V{C}_{\mathbb{J},n, *}^{-1} (\V{y}_{n, *} - \V{\gamma}_* \V{1}), \mbox{ and } \\
		\mbox{variance: } 	&   s_{\mathbb{J}}(\V{x}_*) = c_{\mathbb{J}}(\V{x}_*, \V{x}_*) - \V{c}_{\mathbb{J},n, *}^T \V{C}_{\mathbb{J},n, *}^{-1} \V{c}_{\mathbb{J},n, *}.
	\end{array}
\end{equation}
That is, the posterior distribution of $f(\V{x}_*)$ is only dependent on one side of the local data, $\mathcal{D}_{n, *}(\V{x}_*)$. If $B(\V{x})=0$ approximates the boundary of $A_j$ around the test location $\V{x}_*$ accurately (at least locally), the one-sided local data, $\mathcal{D}_{n, *}(\V{x}_*)$, will contain only the training data belonging to $A_j$. Therefore, the local GP prediction with the one-sided local data would give an accurate estimate of $g_j$. The resulting model is referred to as the \emph{Jump GP}, because it can accommodate discontinuity or jump in the underlying regression surface. The accuracy of the Jump GP model depends on several model parameters: the mean parameters, the covariance parameters and the parameters $\V{\beta}$ for the boundary model $B(\V{x}) = 0$. Let $\V{\theta}$ represent a set of the mean parameters $\gamma_*$ and $\gamma_o$ and the covariance parameters for $c_*$ and $c_o$. In the next section, we will explain the maximum likelihood approach to estimate the model parameters.

One potential issue with the one-sided local estimate is that it could be less accurate than the full local GP estimate when the test location $\V{x}_*$ is interior of one region $A_j$. Given an interior test location, the local data $\mathcal{D}_{n}$ would contain only training data from one region. For such case, the local GP estimate with only one-sided data would have a higher posterior variance than the one with the full local data. because the one-sided estimate uses less data. Therefore, a better approach is to choose one in between the two estimates depending on a test location. Certainly, we do not know whether a test location is near the boundary of $A_j$ or interior of $A_j$. We propose to take both the full local GP estimate and the one-sided local GP estimate, and choose one that better gives the marginal likelihood value. Let $\V{\delta}$ denote a set of the mean parameter $\gamma$ and the parameters of the covariance model $c(\cdot, \cdot)$ for the full GP model \eqref{eq:localGP} . The negative log likelihood of the full local GP model would be 
\begin{equation} \label{eq:loglike_org}
	\begin{split}
		L(\V{\delta}) = \frac{1}{2} \log |\V{C}_{n}| & + \frac{1}{2} \V{y}_{n}^T \V{C}_{n}^{-1} (\V{y}_{n} - \gamma \V{1}),
	\end{split}
\end{equation}
and the corresponding likelihood term $L_{\mathbb{J}}(\V{\theta}, \V{\beta})$ for the one-sided model can be found in \ref{eq:loglike}. Let $\V{\delta}_f$ denote the maximum likelihood estimate (MLE) of $\V{\delta}$ for the likelihood \eqref{eq:loglike_org}, and let $\V{\theta}_{\mathbb{J}}$ and $\V{\beta}_{\mathbb{J}}$ represent the MLE for the model \ref{eq:loglike}. If $L(\V{\delta}_f) < L(\V{\theta}_{\mathbb{J}}, \V{\beta}_{\mathbb{J}})$, we choose the full local estimate. Otherwise, we choose the one-sided estimate as the final estimate of $f$. 

\section{Maximum Likelihood Estimation of $\V{\beta}$ and $\V{\theta}$} \label{sec:estimation}

The proposed jump GP model comes with two sets of parameters: a set including the parameters $\V{\theta}$ of a stationary covariance function $c(\cdot, \cdot)$ and the parameters $\V{\beta}$ of the boundary representation $B(\V{x})=0$. We propose to choose the two hyperparameters using a likelihood maximization approach. To start with, we first write the negative log likelihood function of the two parameters given the local data $\mathcal{D}_n$ can be derived from the joint pdf of the distribution \eqref{eq:jointpdf} as 
\begin{equation} \label{eq:loglike}
	\begin{split}
		L_{\mathbb{J}}(\V{\theta}, \V{\beta}) = \frac{1}{2} \log |\V{C}_{\mathbb{J},n, *}| & + \frac{1}{2} (\V{y}_{n,*} - \gamma_*\V{1})^T \V{C}_{\mathbb{J},n, *}^{-1} (\V{y}_{n,*} - \gamma_*\V{1}) \\
		& + \frac{1}{2} \log |\V{C}_{\mathbb{J},n, o}| + \frac{1}{2} (\V{y}_{n,o} - \gamma_0 \V{1})^T \V{C}_{\mathbb{J},n, o}^{-1} (\V{y}_{n,o} - \gamma_o \V{1}).
	\end{split}
\end{equation}
Normally, the negative log likelihood function is minimized for the parameter estimation by a gradient algorithm such as the Quasi Newton methods \citep{byrd1995limited}. This approach would not work mainly to the non-differentiability of $\mathbb{J}(\cdot, \cdot)$ with respect to the parameters $\V{\beta}$.  Getting the subgradient with respect to $\V{\beta}$ is not straightforward as well, which erases out the possibility of using the subgradient algorithm. Here, we take an analytical approximation of $\mathbb{J}(\cdot, \cdot)$. For the approximation, please note that the term $\frac{B(\V{x})}{|B(\V{x})|}$ can be analytically approximated by a differentiable function $\tanh(\kappa B(\V{x}))$ in that 
\begin{equation*}
	\lim_{\kappa \rightarrow \infty} \tanh(\kappa B(\V{x})) = \frac{B(\V{x})}{|B(\V{x})|}.
\end{equation*}
Therefore, the test function $\mathbb{J}$ is approximated by 
\begin{equation}
	J_{\kappa}(\V{x}, \V{x}') = \frac{1}{2}\left( 1 + \tanh(\kappa B(\V{x})) \tanh(\kappa B(\V{x}')) \right), 
\end{equation}
and 
\begin{equation}
	\lim_{\kappa \rightarrow \infty}  J_{\kappa}(\V{x}, \V{x}') = J(\V{x}, \V{x}') .
\end{equation}
Accordingly, the mean and covariance functions are approximated by
\begin{equation*}
	r_{\mathbb{J}_{\kappa}}(\V{x}) = \mathbb{J}_{\kappa}(\V{x}, \V{x}_*) \gamma_*  + (1-\mathbb{J}(\V{x}, \V{x}_*))\gamma_o,
\end{equation*}
and the covariance function 
\begin{equation}
	c_{\mathbb{J}_{\kappa}}(\V{x}, \V{x}') = c_*(\V{x}, \V{x}') \mathbb{J}_{\kappa}(\V{x}, \V{x}') + c_o(\V{x}, \V{x}') \mathbb{J}_{\kappa}(\V{x}, \V{x}'). 
\end{equation}
With the approximations, we would have the corresponding approximation of the negative log likelihood in the form of
\begin{equation} \label{eq:loglike2}
	\begin{split}
		L_{\mathbb{J}, \kappa}(\V{\theta}, \V{\beta}) = \frac{1}{2} \log |\V{C}_{\mathbb{J}, n}| & + \frac{1}{2} (\V{y}_{n} - \V{\gamma}_{\mathbb{J}, n})^T \V{C}_{\mathbb{J}, n}^{-1} (\V{y}_{n} - \V{\gamma}_{\mathbb{J}, n}),
	\end{split}
\end{equation}
where $\V{C}_{\mathbb{J}, n}$ is a $n \times n$ matrix of the approximate covariance function values $c_{\mathbb{J}_{\kappa}}(\cdot, \cdot)$ evaluated for all pairs of the training inputs in $\mathcal{D}_n$, and $\V{\gamma}_{\mathbb{J}, n}$ is a $n \times 1$ vetor of the approximate mean function $r_{\mathbb{J}_{\kappa}}$ values evaluated for all training inputs in $\mathcal{D}_n$.  We minimize the approximate negative log likelihood with respect to the parameters $\V{\theta}$ and $\V{\beta}$ by a standard gradient descent algorithm.

\begin{figure}[ht!]
	\centering
	\includegraphics[width=0.7\textwidth]{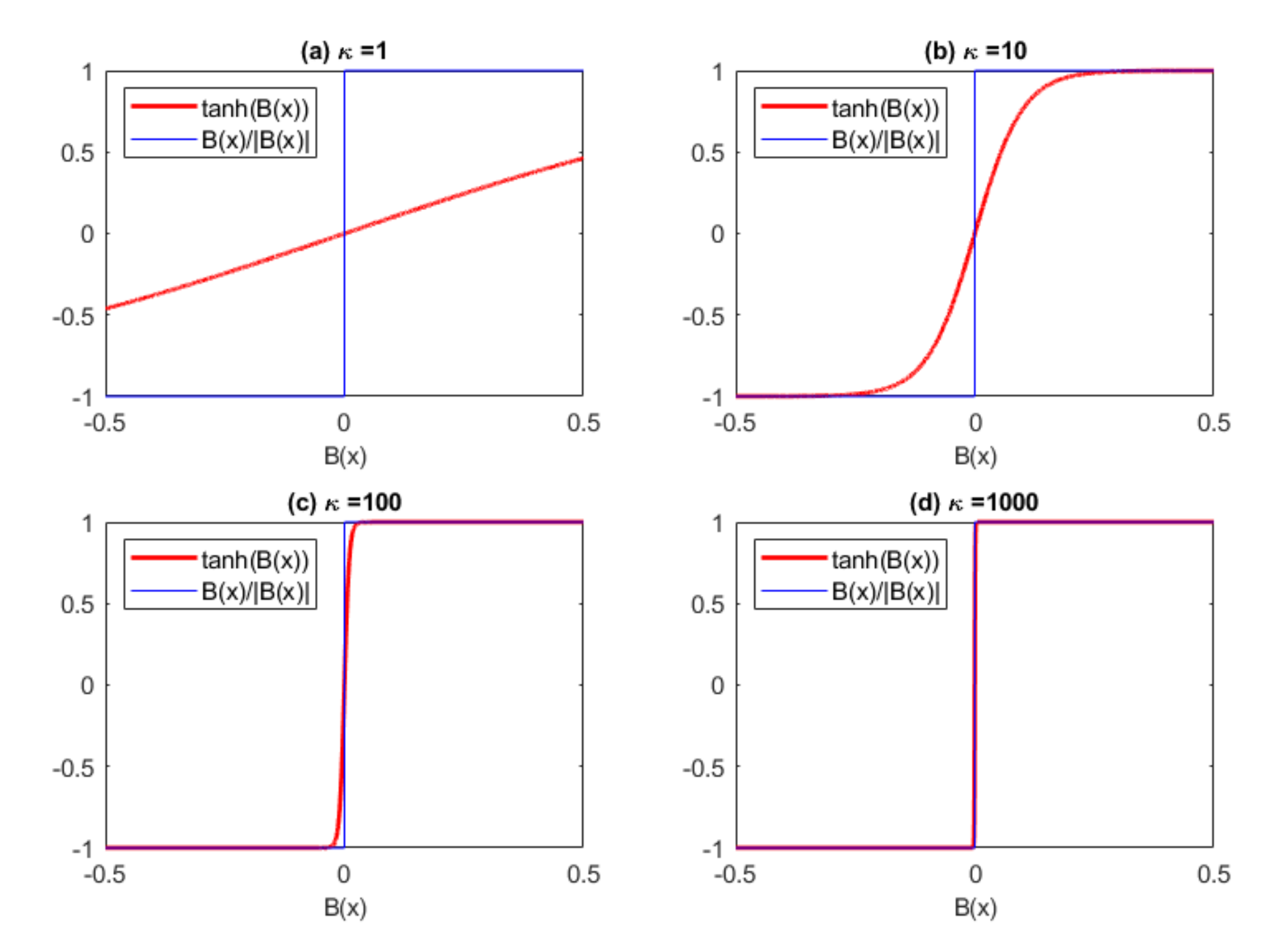}
	\caption{Approximation of $B(\V{x}) / |B(\V{x})|$ by $\tanh(\kappa B(\V{x}))$ with different choices of  $\kappa$.} \label{fig2}
\end{figure}

There are two crucial choices that determine the parameter estimation: the approximation parameter $\V{\kappa}$ and the initial estimates of $\V{\theta}$ and $\V{\beta}$. As $\kappa$ increases, the approximate likelihood converges to the original likelihood. Therefore, theoretically, a larger value of $\kappa$ is desirable. However, if $\kappa$ is too large, the term $\tanh(\kappa B(\V{x}))$ is closer to a staircase function as illustrated in Figure \ref{fig2}, which has zero derivative almost everywhere. Therefore, running an gradient descent algorithm with a large value of $\kappa$ would not change  $\V{\beta}$ at all. Based on our many numerical trials, $\kappa \approx 100$ works very well. The $\tanh(\kappa B(\V{x}))$ with $\kappa=100$ approximates the original term very closely with some degrees of smoothness. 

Another important factor to determine the quality parameter estimate is to provide a good initial estimate. The likelihood function is a non-convex and non-linear function of the parameters, and a gradient descent algorithm only gets a local optimal solution. With a good initial estimate, the algorithm would reach to a better local optimal solution. Particularly, the accurate estimation of the boundary function parameter $\V{\beta}$ is crucial to divide the local data into two sides by the regional boundary, so having a good initial estimate of $\V{\beta}$ is important. One good starting value for $\V{\beta}$ can be achieved using a local linear kernel estimator. According to the existing theory of jump regression \citep{qiu2009jump}, when the test location $\V{x}_*$ is near the boundary of region $A_j$, the normal direction to the local boundary can be estimated as the linear coefficient term of the local linear approximation to an unknown regression surface. If the normal direction is chosen as the direction of $\V{\beta}$, the $\V{B}(\V{x}) = 0$ can be in parallel to the local boundary of $A_j$ near the test location. To describe the idea more precisely, consider a local linear estimator of an unknown regression function $f$ at a test location $\V{x}_*$ using the local data $\mathcal{D}_n$, which is formulated as
\begin{equation}
	\min_{\V{\alpha} \in \mathbb{R}^p, \alpha_0 \in \mathbb{R}} \sum_{i \in I_n(\V{x}_*)} \left[y_i - \alpha_0 - \V{\alpha}^T (\V{x}_i - \V{x}_*)\right]^2. 
\end{equation}
The value of $\V{\alpha}$ that solves the optimization problem is approximately perpendicular to the local boundary of $A_j$ near the test location. Therefore, we use the value as the last $p$ values of $\V{\beta}$. The first value of $\V{\beta}$ corresponds to the linear intercept term, so the initial estimate of the scalar value can be sought using the line search of minimizing the negative log likelihood \eqref{eq:loglike2}, while fixing the other parameters. Once the initial estimate of $\V{\beta}$ is achieved and the initial estimate of $\V{\theta}$ is randomly chosen, one can refine the initial estimates by running a gradient decent algorithm with the negative log likelihood loss. This approach works very well for various simulated scenarios as reported in Section \ref{sec:simulation}. 

\section{Illustration of Jump GP with 2D Toy Examples} \label{sec:simulation}

\begin{figure}[t]
	\centering
	\includegraphics[width=0.6\textwidth]{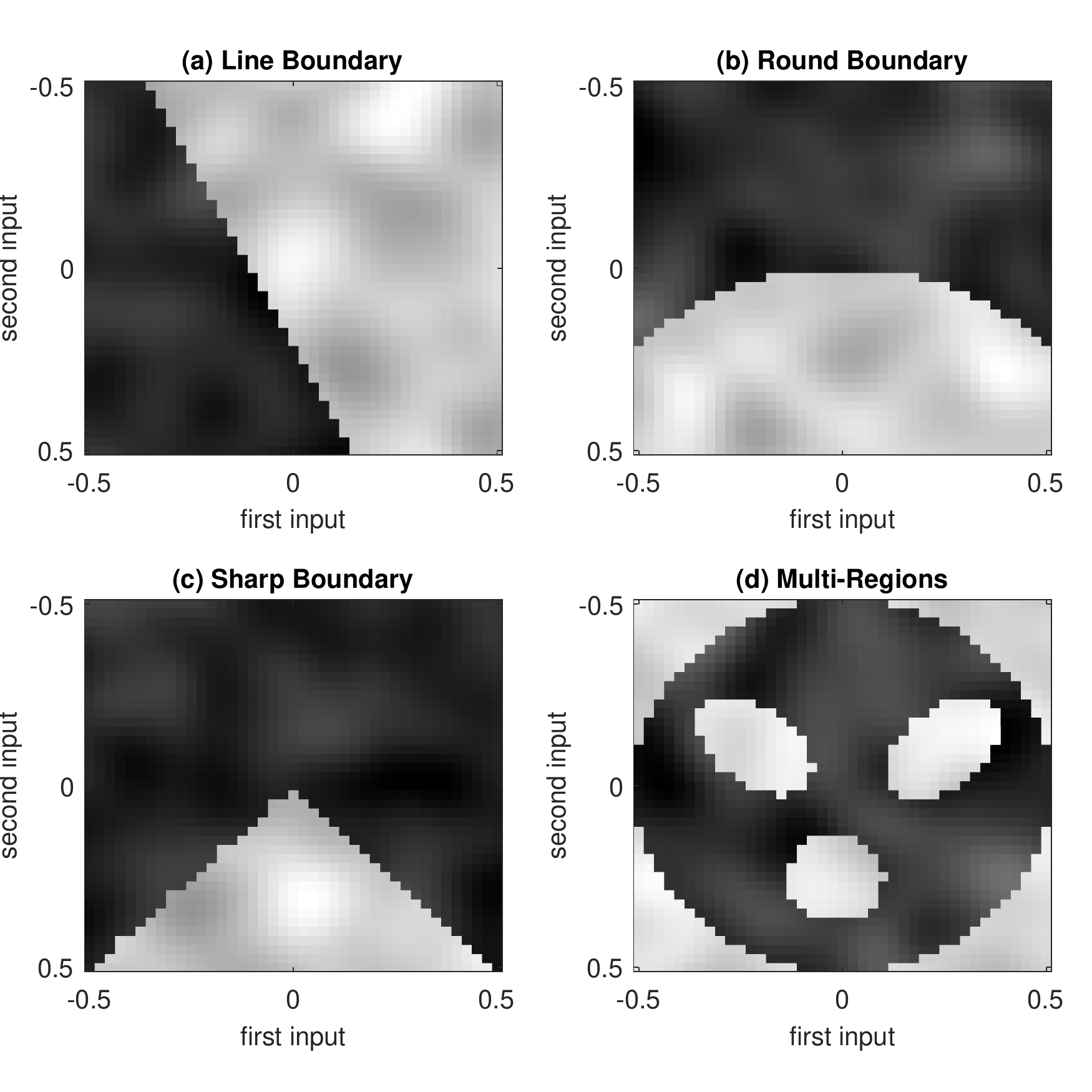}
	\caption{Regression functions of four 2D toy examples. The $x$ and $y$ axes represent the first input and second input values, and the grayscale intensity represents the regression function value for an input location.} \label{fig3}
\end{figure}

This section uses various simulated examples with two dimensional input to illustrate how the proposed jump GP model works with the comparison to the conventional local GP model. Four different regression functions defined on an input domain $[-0.5, 0.5] \times [-0.5, 0.5]$ are selected as shown in Figure \ref{fig3}. For the four cases, the input domains are divided into two or more distinct regions, and an independent realization of Gaussian process with a square exponential covariance function and different constant mean functions (chosen from 0 and 128) is taken for each of the regions. The four cases have different characteristics in the regional boundaries. For the first two cases, the regional boundaries are smooth. In Figure \ref{fig3} (a), the regional boundary is linear, while the boundary is curvy and smooth in Figure \ref{fig3} (b). In Figure \ref{fig3} (c), the regional boundary contains one sharp corner. Please note that our proposed approach is based on the locally linear approximation to the regional boundary, which works well for smooth regional boundaries such as the first two cases but is not ideal for the sharp boundary such as the third case. Lastly, the case with Figure \ref{fig3} (d) has multiple  regions and boundaries.

For each case, the training set of 500 noisy observations is randomly drawn. Each training input $\V{x}_i$ is randomly drawn from the uniform distribution over the input domain. The noisy response variable is taken as the ground truth response plus Gaussian noise with zero mean and variance $\sigma^2$. We varied $\sigma^2$ over $\{1, 4, 9\}$, which correspond to the signal-to-noise ratio, 14 decibel (dB), 8 dB and 4.4 dB respectively. The last case can be considered as a very noisy scenario. The test set consists of 1,681 ground truth responses uniformly spread over the grid locations in the input domain.

\begin{figure}[ht!]
	\centering
	\includegraphics[width=\textwidth]{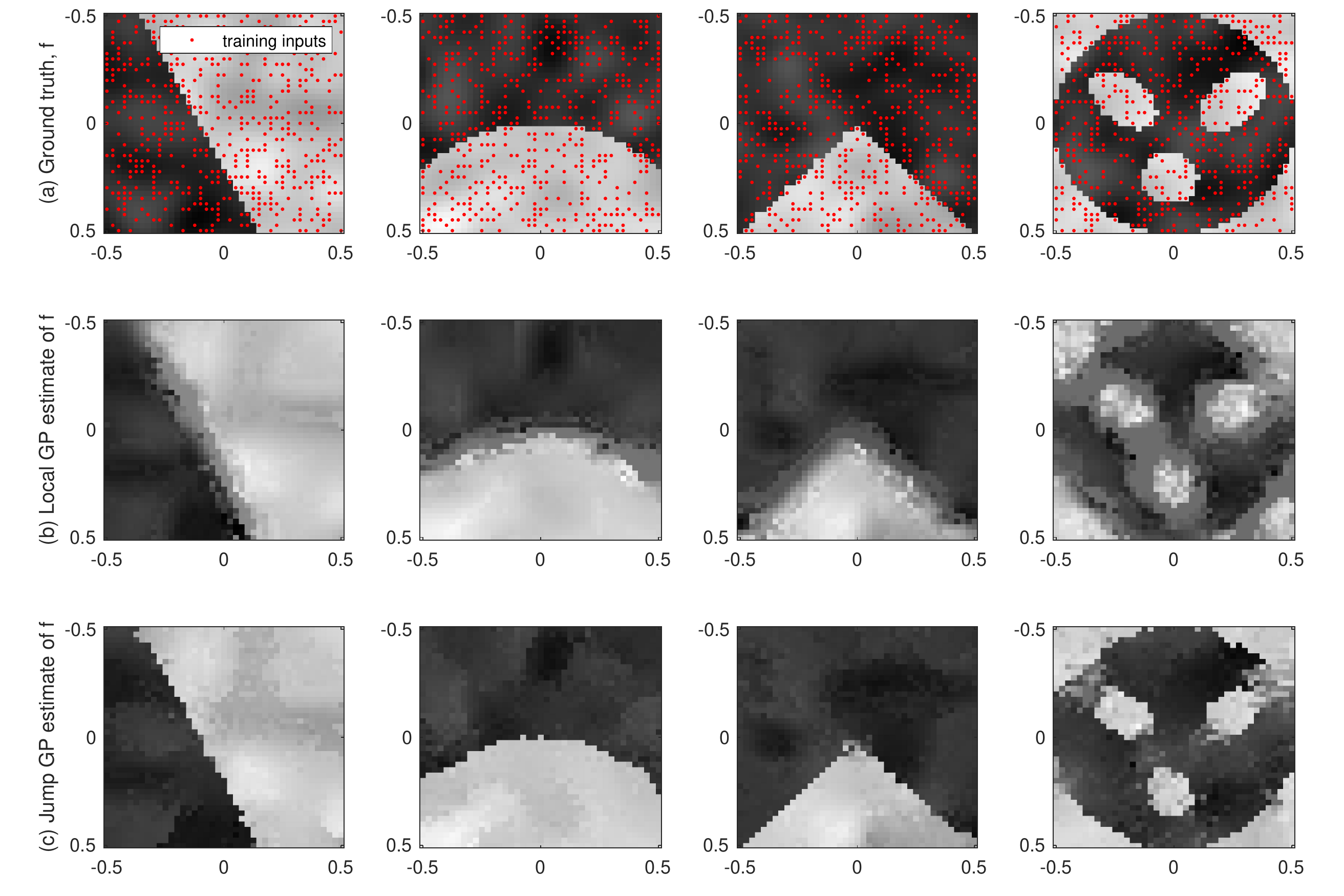}
	\caption{Estimates of $f$ from the local GP and the proposed jump GP. In row (a), the whole simulated non-noisy data are shown with the input locations of the 500 training points annotated as red dots. Rows (b) and (c) show the local GP and jump GP estimates at 1,681 grid locations over $[-0.5, 0.5] \times [-0.5, 0.5]$. Here we used the setting $k=25$ and $\sigma^2 = 4$.} \label{fig4}
\end{figure}

\begin{figure}[ht!]
	\centering
	\includegraphics[width=0.6\textwidth]{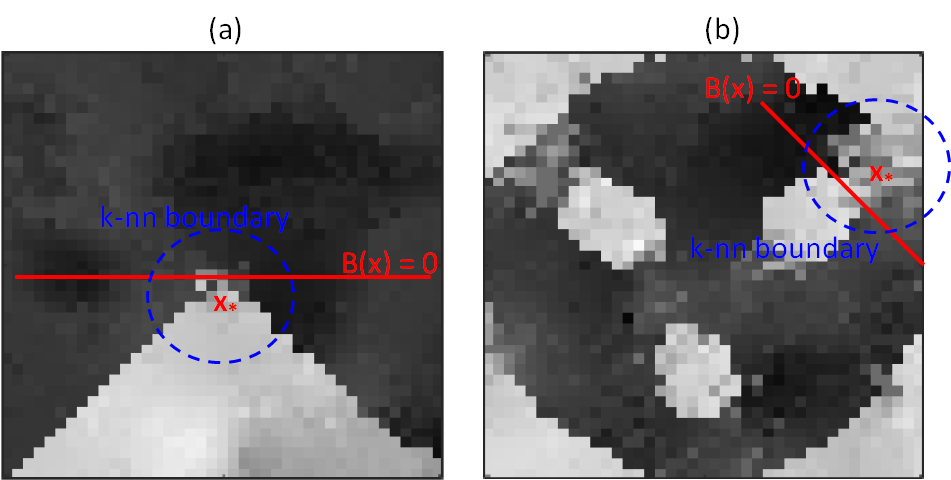}
	\caption{Cases when the proposed approach is less accurate. For those two cases, heterogeneous regression regions in the local neighborhood (surrounded by blue dotted line) cannot be linear separable. } \label{fig6}
\end{figure}

We use the training set to fit the proposed jump GP model and the conventional local GP model. For both of the models, we choose local data using the $k$-nearest neighborhood. We tried different $k$ values, $k \in \{25, 35, 50\}$. The fitted models are evaluated at the test sites, i.e. 1,681 grid locations over $[-0.5, 0.5] \times [-0.5, 0.5]$. Figure \ref{fig4} shows the estimates from the two GP models for the four test regression functions shown in Figure \ref{fig3}. The two estimates are similar to each other when the test inputs locate interior to homogeneous regions. However, they significantly differ on the test locations near the region boundaries. The jump GP estimates are visually more consistent with the ground truth values of $f$ around the boundaries. In particular, the regression estimates of the proposed approach around regional boundaries are very close to the ground truth for the first two test functions that have very smooth regional boundaries. The regression estimates of the proposed approach are less accurate for the last two test functions. For the last two cases, some local data cannot be perfectly divided into heterogeneous regions by a linear boundary. For example, the regional boundary of the third case is triangular with one very sharp tip, around which a linear boundary does not split the local data effectively as illustrated in Figure \ref{fig6} (a). For another example, for the last case, the jump GP estimates around some areas of the darker region sandwiched by two light intensity regions are not accurate, because the local data around the darker areas would contain the data from the two lighter regions, and they are not separable by one linear boundary as illustrated in Figure \ref{fig6} (b). We believe this issue can be addressed by two possible approaches: reducing the neighborhood size $k$ or adopting a quadratic boundary model. Reducing $k$ further could increase the variance of the prediction. We believe using a quadratic boundary model would be a better solution. 

To confirm this visual evaluation, we compare the estimates from the jump GP and the local GP near region boundaries (i.e. test locations within distance 0.05 from the nearest region boundary) to the ground truth response values in terms of two performance metrics: the mean absolute prediction error (MPE) and the mean standardized prediction error (MSPE),
\begin{equation}
	\begin{split}
		& \mbox{MAPE} = \frac{1}{N_t} \sum_{t=1}^{N_t} |\hat{y}_t - y_t|, \mbox{ and } \quad \mbox{MSPE} =  \frac{1}{N_t} \sum_{t=1}^{N_t} \frac{|\hat{y}_t - y_t|}{ s_t },
	\end{split}
\end{equation}
where $N_t$ is the number of the test sites, $\hat{y}_t$ and $y_t$ are the posterior predictive mean estimate and the true value of $f$ at the $t$th test site respectively, and $s_t$ is the posterior predictive standard deviation estimate at the $t$th test site. The MAPE measures the overall error of the posterior mean estimates. The MSPE averages the ratios of the mean errors relative to the corresponding posterior standard deviation estimates. If the MSPE value is $r$, this implies that the true response values are within (posterior mean estimate) $\pm r$ (posterior standard deviation) on average, so the second metric quantifies how the posterior predictive distributions are well fitted to the test data.

\begin{figure}[ht!]
	\centering
	\includegraphics[width=0.8\textwidth]{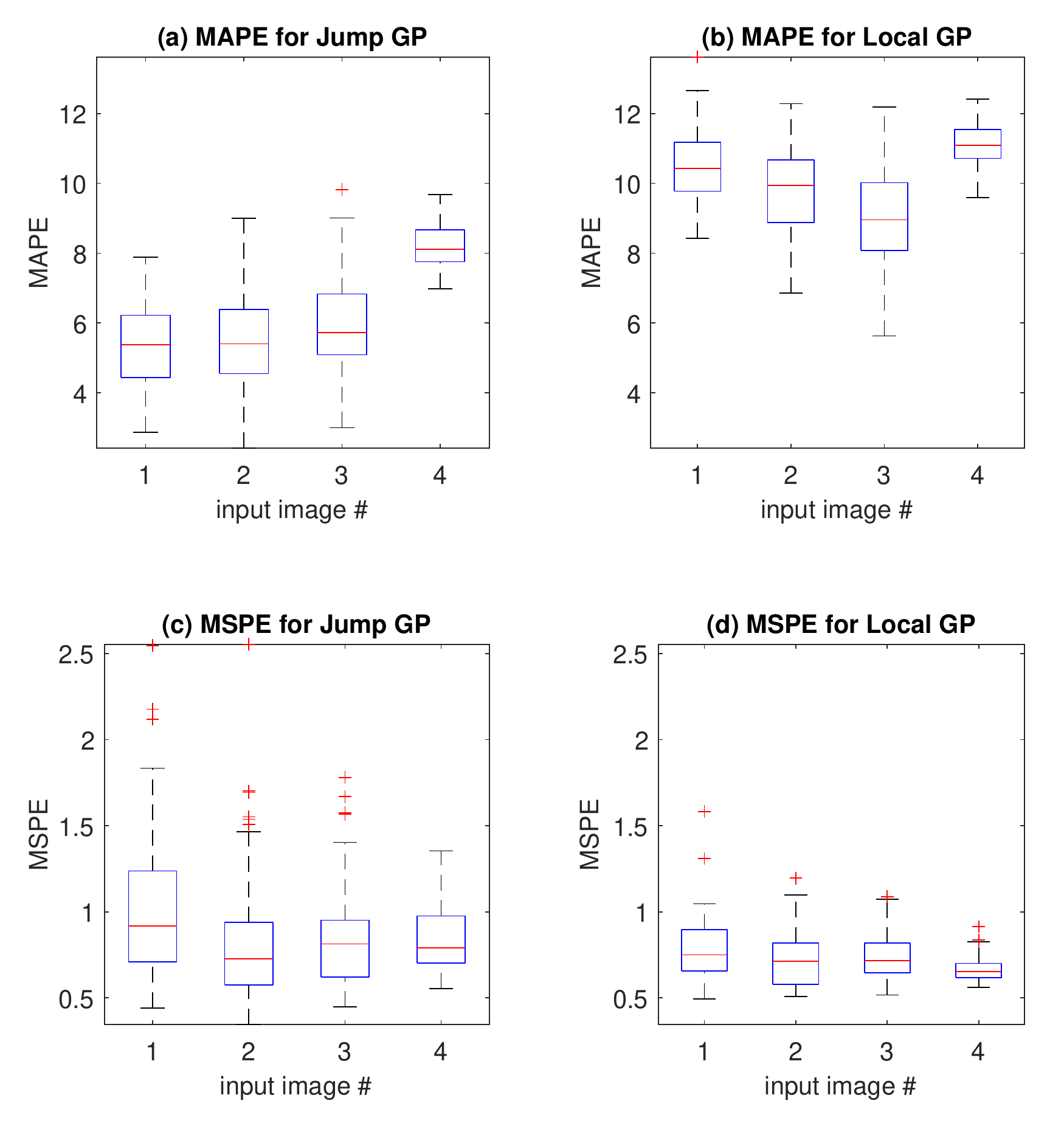}
	\caption{MAPE and MSPE performance of the local GP and the jump GP estimates for the four simulated regression functions shown in Figure \ref{fig3}. In each of the box plots shown above, the red center marks represent the median of the metrics over 75 experiments (three different levels of $\sigma^2$ and 25 replicated experiment per each level). The top and bottom sides of the blue box represent the 25th and 75th percentiles respectively, while the upper and lower black bars represent the median $\pm$ 1.5 (75th percentile - 25th percentile). Red crosses represent the outliers outside the upper and lower bars.} \label{fig5}
\end{figure}

Figure \ref{fig5} shows the summary statistics of the performance metrics over different settings of $\sigma^2$ and 25 replicated experiments per each setting. The plots only show the results with $k=25$, since the results for the other values $k$ do not differ much.  The MAPE metrics of the proposed jump GP are about twice lower than those of the local GP estimates. The MSPE metrics are comparable. The proposed jump GP models tends to have higher variations in the MSPE metrics. 

\section{Conclusion} \label{sec:conc}
We proposed a jump Gaussian process model that accurately estimates an unknown piecewise continuous regression function, where different regression functions are used to model different regions of data. The proposed approach uses the local data neighboring to a test location to make a prediction of the unknown regression function value at the test location. Unlike the conventional local GP approach, the proposed approach identifies a potential regional boundary crossing the local neighborhood and split the local data by the boundary into two sides so as to use only one side of the local data relevant to the regression estimate at the test location. The local data split idea has two major advantages over the existing piecewise Gaussian process models that partition the entire input domain globally in pieces and explicitly model independent regression functions for different areas of the input domain. First, the local split is simpler than the global split, because smooth and complex regional boundaries can be locally approximated by simple linear boundaries so the local data split can be performed by simple boundary models. Second, due to the simplicity, the model estimation of the proposed GP approach is simpler and faster than the approaches based on the global domain split. We showed the two advantages using various simulated scenarios. The proposed local data split by linear boundaries can be extended to the split by higher order polynomial boundaries for handling more complex cases.

\vskip 0.2in
\bibliographystyle{agsm}
\bibliography{jumpgp}

\end{document}